\documentclass{iaus}
\usepackage{graphicx}

\title[ ] 
{Analytic approximations for transit light curve observables
and uncertainties}

\author[J. A. Carter et al.]{Joshua A.~Carter$^1$,
Jennifer C.~Yee$^2$,
Jason Eastman$^2$,\\
B.~Scott Gaudi$^2$,
Joshua N.~Winn$^1$}

\affiliation{$^1$Department of Physics, and Kavli Institute for
  Astrophysics and Space Research, Massachusetts Institute of
  Technology, Cambridge, MA 02139  \\[\affilskip]
$^2$Department of Astronomy, Ohio State University, 140
  W.~18th Ave., Columbus, OH 43210}

\pubyear{2008}
\volume{253}  
\pagerange{?--?}
\jname{Transiting Planets}
\editors{eds. F. Pont et al.}
\begin{document}

\newcommand{\TEEXACT}{T_e}
\newcommand{\TCREXACT}{T_{\times}}
\newcommand{\R}{r}
\newcommand{\PR}{p}
\newcommand{\SE}{S_e}
\newcommand{\AREA}{A}
\newcommand{\PI}{\Pi}
\newcommand{\THETAP}{\tilde{\theta}}
\newcommand{\TCR}{T}
\newcommand{\DEPTH}{\delta}
\newcommand{\FO}{f_0}
\newcommand{\TE}{\tau}
\newcommand{\TF}{T_f}
\newcommand{\TOUT}{T_{{\rm out}}}
\newcommand{\TTRAN}{T_{{\rm transit}}}
\newcommand{\TTOTAL}{T_{{\rm tot}}}
\newcommand{\RHO}{\theta}
\newcommand{\ETA}{\eta}
\newcommand{\TCENTER}{t_c}
\newcommand{\ODD}{\Pi}
\newcommand{\BETA}{\beta}
\newcommand{\TAU}{\tau_0}
\newcommand{\FE}{F^e}
\newcommand{\LE}{\lambda^e}
\newcommand{\FL}{F^l}

\maketitle

\begin{abstract}
  The light curve of an exoplanetary transit can be used to estimate
  the planetary radius and other parameters of interest. Because
  accurate parameter estimation is a non-analytic and computationally
  intensive problem, it is often useful to have analytic
  approximations for the parameters as well as their uncertainties and
  covariances.  Here we give such formulas, for the case of an
  exoplanet transiting a star with a uniform brightness distribution.
  When limb darkening is significant, our parameter sets are still
  useful, although our analytic formulas underpredict the covariances
  and uncertainties.
\end{abstract}

In general, the parameters of a transiting system and their
uncertainties must be estimated from the photometric data using
numerical methods.  However, even when numerical algorithms are
required for precise answers, it is often useful to have analytic
approximations for the parameters as well as their uncertainties and
covariances.  Analytic approximations can be useful for planning
observations, understanding the parameter degeneracies inherent in the
model, calculating order-of-magnitude estimates of the observability
of subtle transit effects, and designing low-correlation parameter
sets that will speed the convergence of optimization algorithms.

\begin{figure}[b]
\begin{center}
 \includegraphics[width=3.0in]{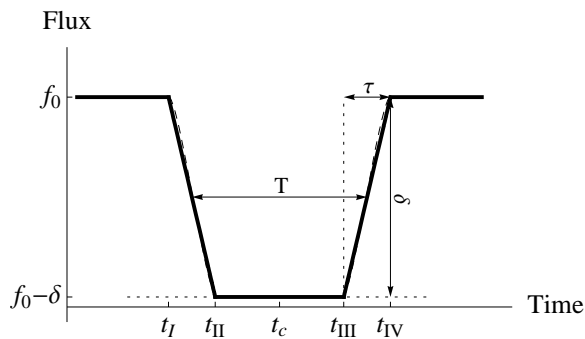} 
 \caption{Linear model for a transit light curve with negligible limb darkening.  Dashed line shows the physical light curve approximated by the linear model. }
   \label{fig:model}
\end{center}
\end{figure}

To perform analytic calculations, we use a piecewise-linear
approximation to the exact transit light curve, parameterized by
$\{\TCENTER, \TE, \TCR, \DEPTH, \FO\}$ as illustrated in
Figure~\ref{fig:model} (see also Seager \& Mallen-Ornelas~2003).
The parameters are the midtransit time ($\TCENTER$), the
out-of-transit flux ($\FO$), the flux decrement during the full phase
of the transit ($\DEPTH$), the duration of ingress or egress ($\TE$),
and the duration between the midpoint of ingress and the midpoint of
egress ($\TCR$). For exact definitions of these parameters in terms of
physical parameters (the impact parameter, scaled semimajor axis, and
radius ratio), see \cite[Carter et al. (2008)]{carter08}.

Given the parameters $\{\TCENTER, \TE, \TCR, \DEPTH, \FO\}$, it is
possible to estimate the covariance matrix of the parameters that
could be derived from observations of a light curve with some
specified number of points $N$ and photometric errors $\sigma$ (see,
e.g., Gould~2003).  The accuracy of the results depends on the
validity of the piecewise-linear model, and the assumptions that the
photometric data points are evenly and densely spaced in time with
independent Gaussian errors. Applying this technique (often referred
to as Fisher information analysis), we obtain some simple and
reasonably accurate expressions for the case in which the planet is
small, the out-of-transit flux is known precisely, and limb darkening
is negligible:
\begin{eqnarray}
	\sigma_{\TCENTER} & = & Q^{-1} \TCR \sqrt{\TE/2 \TCR},  \nonumber\\
	\sigma_{\TE}      & \approx & Q^{-1} \TCR \sqrt{6 \TE/\TCR},  \nonumber\\
	\sigma_{\TCR}     & \approx & Q^{-1} \TCR \sqrt{2 \TE/\TCR},  \nonumber \\
	\sigma_{\DEPTH}   & \approx & Q^{-1} \DEPTH. \nonumber
\end{eqnarray}
In these expressions, $Q \equiv \sqrt{N} \delta/\sigma$, 
the total signal-to-noise ratio of the transit in the
small-planet limit.  Using standard covariance propagation, we can
estimate variances in other observables such as the mean stellar
density in this same limit:
\begin{eqnarray}
	\sigma_{\rho_{\star}} & = & Q^{-1}\rho_{\star}  \sqrt{3 \TCR/8 \TE}  \nonumber
\end{eqnarray}
For the complete covariance matrix, including the more general case in
which the planet is not necessarly small and the out-of-transit flux
is not known precisely, see \cite{carter08}. That work also gives
analytic formulas for other interesting ``derived'' quantities besides
$\rho_\star$, and investigates the degree to which these formulas
underpredict the errors and covariances when limb-darkening is
appreciable. Complementary to this work is the numerical Fisher
analysis for the case of a limb-darkened star recently presented by
P\'al~(2008).

Not only do the parameters $\{\TCENTER, \TE, \TCR, \DEPTH, \FO\}$
yield simple analytic expressions for their uncertainties and
covariances; they are also weakly correlated over the majority of the
allowed parameter space.  As such these parameters provide a simple
analytic framework for weighing the statistical merits of other
parameter choices.  For numerical fitting codes, a desirable parameter
set has low mutual correlations and easily specified {\it a priori}
likelihood functions.  For this purpose we advocate a parameter set
consisting of the midtransit time, the out-of-transit flux, the ratio
of planetary to stellar radii ($R_{p}/R_{\star}$), the normalized
impact parameter, and the duration between the midpoint of ingress and
the midpoint of egress ($\TCR$).  Refer to \cite{carter08} for
analysis of this parameter set and additional parameter sets that are
essentially uncorrelated.

\acknowledgements

We thank Philip Nutzman, Sara Seager and Paul Joss for providing
helpful comments.


\begin{thebibliography}{}

\bibitem[Carter \etal\ (2008)]{carter08}
Carter, J.~A., Yee, 
J.~C., Eastman, J., Gaudi, B.~S., 
\& Winn, J.~N.\ 2008, ArXiv e-prints, 805, arXiv:0805.0238 

\bibitem[Gould (2003)]{gould03}
 Gould, A.\ 2003, ArXiv Astrophysics e-prints, arXiv:astro-ph/0310577
 
 \bibitem[P{\'a}l (2008)]{pal08} P{\'a}l, A.\ 2008, ArXiv 
e-prints, 805, arXiv:0805.2157 
 
 \bibitem[Seager \& Mall{\'e}n-Ornelas (2003)]{seager03} Seager, S., \& Mall{\'e}n-Ornelas, G.\ 2003, \textit{ApJ}, 585, 1038 

\end{thebibliography}
\end{document}